\documentclass{aastex}
\usepackage{spr-astr-addons}
\usepackage[latin1]{inputenc}
\usepackage{url}
\usepackage{graphics}
\usepackage{epsfig}
\usepackage{longtable}
\RequirePackage{color}
\begin{document}
\title{Statistical Study of Observed and Intrinsic Durations among \textit{BATSE} and \textit{Swift/BAT} GRBs}
%\slugcomment{Not to appear in Nonlearned J., 45.}
%% Running heads
%\shorttitle{Short article title} \shortauthors{Authors et al.}
%\author{H.~Zitouni\altaffilmark{1}}
\author{H.~Zitouni}
\affil{ LPTEAM, Facult\'{e} des sciences et de technologie,
Universit\'{e} Dr Yahia Fares, P\^{o}le urbain, M\'{e}d\'{e}a,
Algeria.} \email{zitouni.hannachi@gmail.com}\and

\author{N.~Guessoum} \affil{Department of Physics, College of Arts
\& Sciences, American University of Sharjah, UAE.}\email{
nguessoum@aus.edu} \and
\author{W.~J.~Azzam} \affil{Department of Physics, College of
Science, University of Bahrain,
Bahrain.}\email{wjazzam@sci.uob.bh} \and
\author{R.~Mochkovitch} \affil{UPMC-CNRS, UMR7095, Institut d'Astrophysique de Paris, \mbox{F-75014}, Paris, France.}\email{mochko@iap.fr}

 %\altaffiltext{1}{Faculté des sciences, Universit\'{e} Dr Yahia Fares, Pole urbain, M\'{e}d\'{e}, Algeria.}
 %\altaffiltext{2}{Department of Physics, College of Arts \& Sciences, American University of Sharjah, UAE.}
 %\altaffiltext{3}{Department of Physics, College of Science, University of Bahrain, Bahrain.}

\begin{abstract}

Studies of \textit{BATSE} bursts \citep{kouveliotou:93} have
resulted in the widespread adoption of a two-group categorization:
long bursts (those with durations $\geq 2$ seconds) and short
bursts (those with durations $\leq 2$ seconds). This
categorization, one must recall, used the observed $T_{90}$ time
durations for bursts (during which 90\% of a burst's fluence is
measured).

 In this work, we have explored two ideas: 1) a statistical
search for a possible third, intermediate category of bursts
(between the ``short" and the ``long" ones) among 2041
\textit{BATSE} GRBs and 757 \textit{Swift/BAT} ones; 2) a study of
bursts' intrinsic durations, where durations in the bursts'
reference frames (instead of the observed durations) are
considered; for this, 248 \textit{Swift/BAT} bursts that have
redshift measurements were statistically analyzed for the same
categorization goal.

We first use a Monte Carlo method to determine the proper binning
of each GRB, considering that bursts come with different
uncertainties on their durations. Then, using the method of
minimization of chi-square $\chi^2$, we search for the best fit of
the normalized frequency distributions
$\frac{1}{N_0}\frac{dN}{d\ln{T}}$ of durations; this allows us to
compare fits with two groups (``short" and ``long") with fits with
three groups (``short", ``long", and ``intermediate").

Our results indicate that the distributions of observed durations
 are better fitted by three groups than two groups for \textit{Swift/BAT} data; interestingly, the
``intermediate" group appears rather clearly for both observed and
intrinsic durations. For BATSE data, the statistical test does not
prefer three groups over two.

We discuss the results, their possible underlying causes, and
reasonable interpretations.

\end{abstract}


\section{Introduction}
Gamma-ray bursts (GRBs) have durations ranging from 0.001 to 1000
seconds, with substantial variations in their time profiles,
indicating that the sources are very compact ($c\Delta t~<~3000$
km). The energy of photons emitted in the band [1 keV to 10 MeV]
and the total isotropic energy released in a given event are both
huge ($10^{51} - 10^{54}$ ergs).

After the discovery of the
afterglow of GRB970228 with BeppoSAX, the extragalactic origin of
gamma-ray bursts was confirmed \citep{{costa:97},{paradijs:97}}.
The astounding characteristics that those measurements implied
gave strong motivation for researchers to study the phenomenon and
its physical features, as it was evidently linked to the
universe's most distant regions. Indeed, the farthest GRB observed
until now is GRB090429B, which has a redshift z = 9.4, as
determined by photometric techniques \citep{cucchiara:11}.

After a large number of bursts had been observed, statistical
studies were conducted to classify GRBs, in the aim of finding
correlations between their physical characteristics or finding
links to other better known phenomena such as supernovae. One of
these classifications, which we are exploring in this work, is the
distribution of GRBs' observed and intrinsic durations.

The classification of GRBs according to their intrinsic
properties, i.e. estimated in their own rest frames, is very
important to understanding the physics of these phenomena. This
task, however, is possible only after having collected a large
number of bursts with measured redshifts, such that analyses and
statistical tests can be performed. Before enough redshifts became
available, classifications were made only on the basis of observed
quantities such as the duration of the bursts $T^{90}_{obs}$
during which 90\% of the fluence is accumulated.

Using the first BATSE catalog, \cite{kouveliotou:93} and
\cite{mcbreen:94} inferred a bi-modal distribution for the
logarithm of the duration $T^{90}_{obs}$. The classification of
the distribution of the durations of the GRBs detected by BATSE
into three groups was explored by
\citep{{horvath:98},{balastegui:01},{hakkila:00c},{horvath:02},{chattopadhyay:07}}.
A study by \cite{horvath:09} of BeppoSAX GRBs showed a different
distribution than previously found for the BATSE bursts. Further
research on this classification issue was made on Swift/BAT GRBs
and Fermi/GMB GRBs
\citep{{horvath:08},{zhang:08},{lin:08},{qin:13}}. A study of the
distribution of 1003 GRBs observed by BeppoSAX and given by
\cite{{frontera:09b},{frontera:09a}} was performed by
\cite{horvath:09b}. Comparative studies between the distributions
of the burst durations in BATSE and Swift/BAT samples have been
made by \cite{huja:09}.

The distribution of intrinsic durations of GRBs was studied by
\citep{{zhang:08}}, who used the 95 bursts which then had known
redshifts. The authors used statistical tests to determine whether
the distribution of burst durations was better fitted with two or
three groups. They also compared with data from earlier instruments.
One of the goals of our work was to see whether the analysis of the
larger sample that we have (248 Swift bursts with well-determined
redshifts) would give similar results as those of \citep{{zhang:08}}.
We will see that some interesting similarities and differences have appeared.

In presenting our work, we start by explaining our method of
sampling and analysis of the data used. Following that, we apply
our method by comparing a sample of 2041 GRBs observed by BATSE
and a sample of 757 GRBs observed by Swift/BAT. Next, we compare
the data from the large Swift/BAT sample with that from a
sub-sample consisting of 248 GRBs with known redshifts. The third
section is devoted to the study of these three samples by
comparing their classification into two or three groups by
duration, using the $\chi^2$ test as a measure. We conclude with a
discussion and conclusion.

\section{Method}

In this work we are interested in three samples of bursts observed
by BATSE and Swift/BAT. The first sample consists of 2041 GRBs
observed by BATSE and presented in the ``Current BATSE
Catalog" website\footnote{http://www.BATSE.msfc.nasa.gov/BATSE/grb/catalog}.
The catalog gives: the GRB's number; the burst's duration
$T_{90}^{obs}$; and the uncertainty over it, $\Delta
T_{90}^{obs}$. The second sample consists of all the bursts
observed by Swift/BAT until 13.02.2014 for which the duration and
the uncertainty attached to it are determined. This second sample
is composed of 757 GRBs, for which the data are given on the
``Swift Ground Analysis"
website\footnote{http://gcn.gsfc.nasa.gov/swift$\_$gnd\_ana.html}
. The third sample consists of GRBs from among the second sample
that have a known redshift. This third sample is composed of 248
GRBs, the most recent one being GRB140213A.

The durations of the GRBs in the samples vary in the range
$[10^{-3}, 10^{3}]$ seconds. Studying the distribution of bursts
according to their durations requires (in principle) dividing the
intervals into equal-length bins and filling them with bursts of
the corresponding durations. However, the durations vary by six
orders of magnitude, and the observed data follow a log-normal
distribution. Therefore, it is preferable to use logarithmic
binning in order to better study these distributions.

To construct new distributions on the basis of this choice of
bins, we adopt a two-stage method.

First, we simulate/reconstruct each sample by Monte Carlo. Indeed,
the number of GRBs per bin depends on both the latter's width and
the uncertainty over the bursts' durations. To take the
uncertainties over the durations into account, we simulate large
numbers of GRBs with durations randomly drawn from a normal
distribution $f(T)$ centered on each observed duration, $\mu =
T_{90}^{obs}$, and characterized by a standard deviation $\sigma$
corresponding to the given uncertainty  $\Delta T_{90}^{obs}$. The
normal distribution is given by the expression:
\begin{equation}
f(T)=\frac{1}{\sigma\sqrt{2\pi}}\exp{-\frac{(T-\mu)^2}{2\sigma^2}}.
\end{equation}
The burst's duration in each iteration is calculated by using
Monte Carlo method according to the expression:
\begin{equation}
 T_{i}=F^{-1}(Y_{Rn}).
\end{equation}
The Monte Carlo method is used to draw a random number $Y_{Rn}$
from a uniform distribution between [0, 1] satisfying:
\begin{equation}
F(T)=\int_{-\infty}^{T} f(T') dT'.
\end{equation}

The number of iterations depends on the accuracy that we aim for,
which is in practice set by the computational speed of our
machines and codes. In our work, the number of iterations is
$6\times10^4$.

\begin{figure}[h]
             \centering
       \includegraphics[angle=0, width=0.47\textwidth]{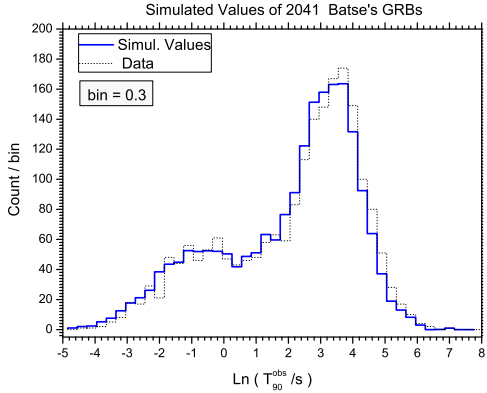}
        \caption{\emph{Comparison between the distribution of observed durations for \textit{BATSE} GRBs and that
        obtained using our Monte Carlo simulation of the same sample. In this case the bin has a width of 0.3}.}
        \label{fig1}
\end{figure}

\begin{figure}[h]
             \centering
       \includegraphics[angle=0, width=0.47\textwidth]{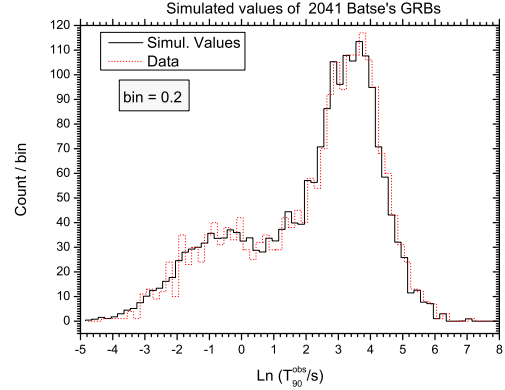}
        \caption{\emph{Comparison between the distribution of observed durations
       for \textit{BATSE} GRBs and that obtained using our Monte Carlo simulation of the same sample. In this case the bin has a width of 0.2.}}
        \label{fig2}
\end{figure}

Through this method, we reconstructed the first sample, which
consists of 2041 \textit{BATSE} GRBs that were observed over 9
years. Furthermore, by increasing the number of simulation values,
this method can smooth the distribution of bursts' durations. In
Fig.\ref{fig1} we present the result of this method for the BATSE
sample, with a logarithmic bin of 0.3. In Fig.\ref{fig2}, we
present the same results with a bin of 0.2 in the aim of showing
the effect of the choice of the width of this parameter on the
distribution. Comparing (Fig.\ref{fig1}) and Fig.\ref{fig2}, we
note that the choice of the width of the bin is important for the
study of each distribution.

Moreover, the different total number of bursts $N_0$ in each of
the three samples that we will be analyzing will not allow us to
make direct comparisons between the different distributions. To
address this problem, we use a method of standardization to bring
all the distributions to the same scale:

\begin{equation}\label{eq1}
g(T)=\frac{1}{N_0}\frac{dN}{d(\ln{T})}.
\end{equation}

This second step in our method thus produces a function $g(T)$
which represents a probability density.

\begin{figure}[h]
             \centering
       \includegraphics[angle=0, width=0.47\textwidth]{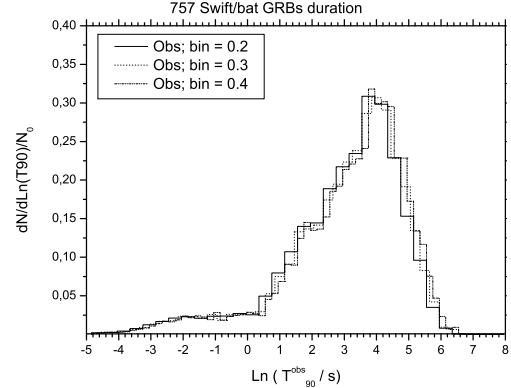}
        \caption{\emph{Normalized distributions of observed durations for \textit{Swift/BAT} bursts and for
        the sample obtained using our Monte Carlo simulation, for bin widths = 0.2, 0.3 and 0.4.}}
        \label{fig3}
\end{figure}

We then use this expression for the 757 \textit{Swift/BAT} GRBs
of our second sample taking bin widths of 0.2, 0.3, and 0.4. We
note from Fig.\ref{fig3} that the resulting distribution profile
is independent of the choice of the bin.

\section{Comparison between \textit{BATSE} GRBs and \textit{Swift/BAT} GRBs}

  In Fig.\ref{fig3} one can readily note that \textit{Swift/BAT} detects fewer SGRBs (short bursts)
  compared to \textit{BATSE}. \textit{Swift/BAT} has higher sensitivity and angular resolution than
  \textit{BATSE}, \textit{BeppoSAX}, and \textit{HETE-2}. However, it is less sensitive to SGRBs than \textit{BATSE} (8\% vs. 18\%).
  This is due mainly to its energy range [15-150 keV] since SGRBs have relatively harder
  spectra \citep{{band:06a},{band:06b},{gehrels:07}}. Moreover, \textit{Swift/BAT} GRBs
  show a clear(er) bump in their distribution of durations between 2 and 10 sec compared to
  the \textit{BATSE} data. This same bump also appears in the distribution of durations of 1003 GRBs
  observed by \textit{BeppoSAX} \citep{{frontera:09a},{horvath:09b}}, which also observed few SGRBs (112)
  compared to LGRBs (891).

\section{Study of the Intrinsic Durations}

It is very important to characterize GRBs by their intrinsic
properties, as these relate to the physical processes that produce
the phenomenon. In relation to this, we may mention the
correlations between the isotropic energy $E_{iso}$, peak  energy
of the spectrum $E^s_p=E_p^{obs}\times (1+z)$, the isotropic
luminosity $L_{iso}$, and the total collimated energy $E_{\gamma}$
\citep{{amati:02},{amati:06},{yonetoku:04},{ghirlanda:06},{nava:12},{zitouni:14}}
and others. The study of these relationships was possible only
after the determination of the GRBs' redshifts.

In this section, we are interested in the study of the distribution of
intrinsic \textit{Swift/BAT} GRB durations. Except for
\cite{zhang:08}, this relationship has not been examined before due
to the low number of GRBs with known redshifts. For this purpose, we constructed
a third sample, which consists of 248 GRBs whose redshifts and
durations have been determined. The intrinsic duration, denoted
$T_s^{90}$, represents the duration of the GRB in its inertial
frame and is inferred from the observed duration via the
expression:

 \begin{equation}
    T_{90}^s=\frac{T_{90}^{obs}}{1+z}.
 \end{equation}

In the sample of Swift/BAT GRBs with known redshifts, we must note
that the distribution of redshifts is not the same for short and
long bursts. For SGRBs, the median of earlier samples of the
redshift distribution was 0.4
\citep{{norris:06},{oshaughnessy:08},{horvath:08},{zhang:08}},
while for LGRBs, the median was 2.4
\citep{{bagoly:06},{horvath:08}}. In our sample, the median of the
redshift distribution for short bursts is $0.78\pm 0.70$, while
for long bursts it is $2.13\pm 1.35$. (The difference in the
values is due to our use of a much larger number of GRBs with
redshifts compared to what these authors used.)

This difference between the redshifts of SGRBs and LGRBs tends to
blur the "boundary" between the durations of the two groups, a
"boundary" traditionally thought to be at 2 s. With an average
redshift of 2, "long" bursts with observed durations less than 6
seconds will have an intrinsic duration below 2 seconds, while
short observed bursts of any redshift will remain short in terms
of intrinsic duration.

 \begin{figure}[h]
             \centering
       \includegraphics[angle=0, width=0.47\textwidth]{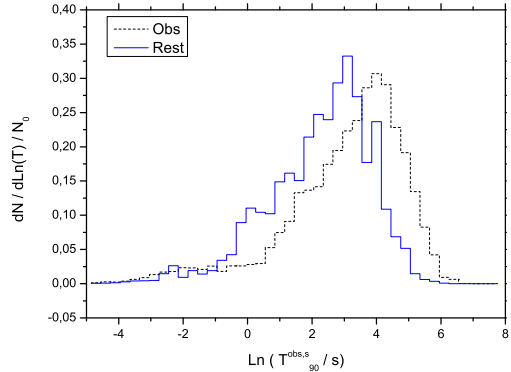}
        \caption{\emph{Comparison between the normalized distribution of observed durations and intrinsic durations for 248 \textit{Swift/BAT} GRBs.}}
        \label{fig3b}
\end{figure}

In Fig.\ref{fig3b} we have represented the two normalized
distributions for observed durations $T^{obs}_{90}$ and intrinsic
durations $T_{90}^{s}$. Note that the distribution of intrinsic
durations is similar to that of the observed durations, with a
general shift to the left of most of the distribution. The area or
boundary that seemed to clearly separate the two populations in
the \textit{BATSE} sample is almost absent in the
\textit{Swift/BAT} sample. Likewise, in the distribution of
intrinsic durations, the clear separation between the short and
the long bursts tends to disappear.

\section{Search for "Intermediate" Duration Group}

The various studies done on the distributions of GRB durations
that we mentioned in the introduction revolved around determining
the groups that the observed durations of GRBs seemed to fall
under in different samples. In the present work, we are interested
in reviewing the groups that GRBs have been categorized under, in
the \textit{BATSE} and the \textit{Swift/BAT} catalogs; in
particular, we are interested in searching for a third,
"intermediate" population of bursts, between the SGRBs and the
LGRBs. For this purpose and for each sample, we perform a
statistical comparison between a given set of data and either two
or three Gaussian distributions.

The first step is to fit a given set of data (\textit{BATSE},
\textit{Swift/BAT}, observed durations, calculated intrinsic
durations) using log-normal functions. The goal is to get the
three parameters used in the Gaussian functions (mean, peak, and
standard deviation values) for a best fit. For a fit with two
groups (or classes), we need six parameters, and for three groups
(or classes) we need nine parameters. We have adopted the
minimization of $\chi^2$ as a way to determining the parameters
for  a best fit in each case
\citep{{baker:84},{hauschild:01},{press1989},{zwillinger2010crc},{saporta2011probabilites},{martin2012statistics}}.

 The $\chi^2$ is defined as:
\begin{equation}
 \chi^2 = \sum_{i=1}^n \frac{(O_i - E_i)^2}{E_i},
\end{equation}
where $O_i$ is the observed value and $E_i$ is the expected value
in the $i$ bin. We use n = 43 bins for the three samples that we
study below. The degree of freedom $\nu = n - k - 1$, where k is
the number of parameters for each model.

The probability density function of the $\chi^2$ distribution with
$\nu$ degrees of freedom is defined in $[0,+\infty]$ as:
\begin{equation}
 f(\chi^2) = \frac{1}{2^{\nu/2}\Gamma(\nu/2)}
 \exp{[-\frac{1}{2}\chi^2}] ~~(\chi^2)^{\frac{\nu}{2}-1}.
\end{equation}

The probability that the observed $\chi^2$  for a correct model
should be less than a critical value, $\chi^2_c$,  with $\nu$
degrees of freedom  is :
\begin{equation}
 P(\chi^2\mid\nu) = \int_0^{\chi^2_c}
f(\chi^2)~d\chi^2,
\end{equation}

Its complement, called "p-value", is the probability that the
observed $\chi^2$ will exceed the critical value $\chi^2_c$
\citep{{saporta2011probabilites},{martin2012statistics}}:
\begin{equation}
\textmd{p-value} = 1-P(\chi^2\mid\nu) = \int_{\chi^2_c}^{+\infty}
f(\chi^2)~d\chi^2.
\end{equation}

 It is defined  by \cite{martin2012statistics} as the smallest level
of significance that would lead to a rejection of the null
hypothesis using the observed sample, or the probability of
getting from the statistical test a value that contradicts the
null hypothesis at least as much as for computed from the sample.
P-values are thus often coupled to a significance level or
$\alpha$, which is also set ahead of time, usually at 0.05, 0.01,
or even smaller \citep{martin2012statistics}.

 We apply our method involving the probability density
distribution functions so that the statistical properties of each
sample are preserved, and the fitting parameters ($\mu$ et
$\sigma$) are generic. We use log-normal functions for these fits,
so the fits can be written as

\begin{equation}
 \frac{1}{N_0}\frac{dN}{d\ln{T}}=\sum_{i=1}^n A_i\exp{-\frac{(ln T-\mu_i)^2}{2\sigma_i^2}},
\end{equation}
where $A_i$ is the amplitude or the maximum height of each group,
$\sigma_i$ is the "width" of the group, $\mu_i$ is the "position" (center)
of the group, and $n$ is the number of groups.

\begin{figure}[h]
             \centering
       \includegraphics[angle=0, width=0.43\textwidth]{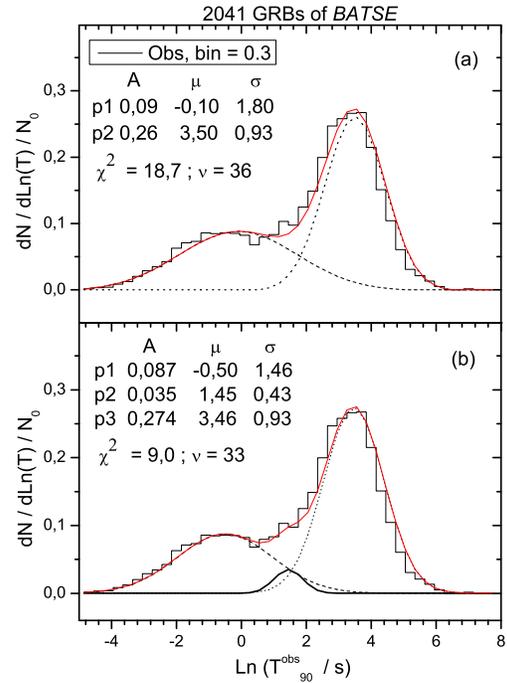}
        \caption{\emph{\textit{BATSE} normalized distribution of GRBs observed durations. The $\chi^2$ is calculated before
        normalisation. $p_1$, $p_2$ and $p_3$ correspond to the
        first, second and third peaks.
        (a): Data  fitted with two lognormal functions. (b): Data fitted with three lognormal
        functions.}}
        \label{fig4}
\end{figure}

In Fig.\ref{fig4}a and Fig.\ref{fig4}b we display the fits of the
distribution of observed durations by considering two or three
groups.

For \textit{BATSE} observed durations, the minimum values for
$\chi^2$ obtained are 18.7  for two groups and 9.0  for three
groups. The parameters of each peak are shown in Fig.\ref{fig4}.

Tab.\ref{tab1} gives the positions of the centers of each group as
obtained in this work, alongside those of \cite{horvath:02}. The
results are in good agreement.
\begin{table}[h]
\centering
\begin{tabular}{lcc}
  \hline
   & $T_{90,m}^{obs}$ (s) &   \\
   & This work & Ref$^a$ \\
   \hline
   \hline
  p1 & $0.99^{+0.11}_{-0.09}$ & $0.78$  \\
  p2 & $33.11^{+0.67}_{-0.65}$ & $34.67$  \\
  \hline
  p1 & $0.61^{+0.06}_{-0.06}$ & $0.56$   \\
  p2 & $4.26^{+0.36}_{-0.33}$ & $4.26$  \\
  p3 & $31.82^{+0.55}_{-0.54}$ & $35.48$  \\
  \hline
\end{tabular}
\caption{Centers of the two- and three-group fits of the
\textit{BATSE} burst durations. (a): \cite{horvath:02};}
 \label{tab1}
\end{table}

\begin{figure}[h]
             \centering
       \includegraphics[angle=0, width=0.47\textwidth]{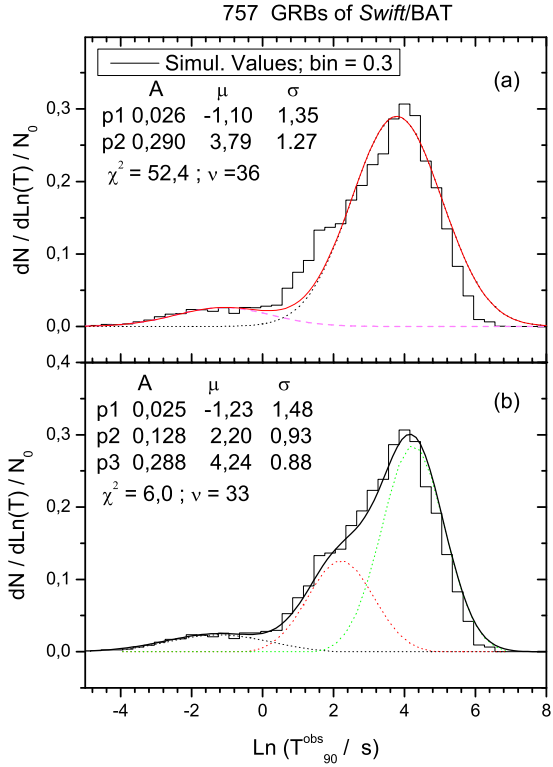}
        \caption{\emph{\textit{Swift/BAT} normalized distribution of GRBs observed durations. The $\chi^2$  is calculated before normalisation. }}
        \label{fig5}
\end{figure}

We apply the same method (minimization of $\chi^2$) for our sample
of Swift/BAT bursts, again fitting the distribution of observed
durations. The minimum values for $\chi^2$ obtained are 6.0
 for three groups and  52.4   for
two groups. The results are shown in Fig.\ref{fig5}a and
Fig.\ref{fig5}b.

Tab.\ref{tab2} gives the positions of the centers of each group,
as obtained in this work, alongside those of \cite{horvath:02}.
The results are in good agreement for the model with three groups.

\begin{table}[h]
\centering
\begin{tabular}{lcccc}
  \hline
  & $T_{90,m}^{obs}$ (s)& & $T_{90,m}^{s}$ (s)& \\
   & This work & Ref.$^b$ & This Work &   Ref$^{c}$ \\
   \hline
   \hline
  p1 & $0.33_{-0.09}^{+0.13}$ &$0.35$ & $0.13^{+0.05}_{-0.03}$ & $0.13$  \\
  p2 & $45_{-3}^{+2}$ & $40.36$ & $16.4^{+1.2}_{-1.1}$ &  $12.30$  \\
  \hline

  p1 &  $0.29_{-0.09}^{+0.13}$ & $0.34$  & $0.11_{-0.03}^{+0.05}$&    \\
  p2 &  $9.06_{-1.23}^{+1.42}$ & $12.79$ & $2.56_{-0.94}^{+1.49}$ &   \\
  p3 &  $69.72^{+4.92}_{-4.59}$ & $79.98$   & $25.0_{-3.6}^{+4.2}$ &   \\
  \hline
\end{tabular}
\caption{Centers of the two- and three-group fits of the
\textit{Swift/BAT} bursts. (b):\cite{horvath:09}; (c):
\cite{zhang:08}}
 \label{tab2}
\end{table}

\begin{figure}[h]
             \centering
       \includegraphics[angle=0, width=0.47\textwidth]{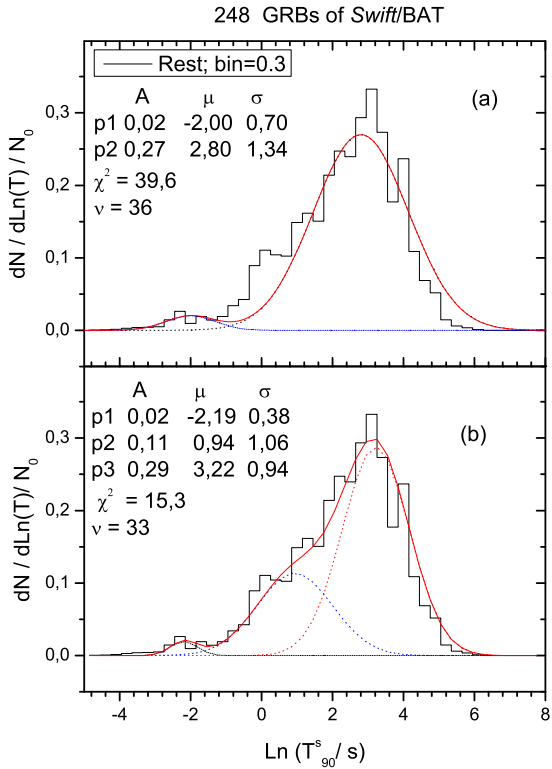}
        \caption{\emph{\textit{Swift/BAT} normalized distribution of GRBs intrinsic durations.  The $\chi^2$  is calculated before normalisation. }}
        \label{fig6}
\end{figure}

The final sample of GRBs studied here consists of 248
\textit{Swift/BAT} GRBs with known redshifts. We applied the same
method, i.e. fitting theoretical lognormal functions to the
intrinsic durations data (i.e. as calculated in the GRBs' source
frames). The minimum values for $\chi^2$ obtained are 39.6  for
two groups and  15.3  for three groups. The main results are
presented in Fig.\ref{fig6}a and Fig.\ref{fig6}b. Tab.\ref{tab2}
(last two columns) gives the positions of the centers of each
group, as obtained in this work (with 248 GRBs), alongside those
of \cite{zhang:08} (with 95 GRBs), for the two-group fits only.

Some authors have used the p-value as a measure of  the goodness
of the fit
\citep{{hauschild:01},{press1989},{press2007numerical}}. If
p-value $> 0.1$, the model can be accepted;  if p-value $< 0.001$,
the model is very likely to be wrong. We do not adopt that kind of
approach; instead of follow the method outlined below.

The second step is to perform a statistical test to compare
between the model of two groups (2g: short and long durations)
with the model of three groups (3g: short, intermediate and long
durations). To perform this test we follow the method used by
\cite{band:97} and \cite{horvath:98}, a method which is well
explained in \cite{balakrishnan1999statistics}.

We calculate $\Delta \chi^2(\nu)~=~\chi^2_{2g}(\nu_1)-
\chi^2_{3g}(\nu_2)$ for each sample of data. The degree of freedom
is $\nu = \nu_1 - \nu_2$. Our null hypothesis here is that the
data is well represented by two groups. We adopt a critical value
of 0.001 for $\alpha$: if $\Delta\chi^2 < \chi^2_{0.001}$ or
(p-value $> 0.001$), then the data is fitted with two groups at a
significance level of 0.001. In all samples, $\nu = 36 - 33 = 3$,
then $\chi^2_{0.001}$ is equal to 16.27. This value is compared to
those obtained ones for each sample of data (\textit{BATSE},
\textit{Swift/BAT}, observed and intrinsic)  presented in the
Table (\ref{tab3}) with their  corresponding p-values.

\begin{table}[h]
\centering
\begin{tabular}{cccc}
  \hline
    & $\Delta\chi^2$ & p-value & Decision \\
    \hline
  BATSE & 9.7  & $2.1~10^{-2}$& H0\\
  Swift/Obs & 46.6 & $4.7~10^{-10}$ & H1 \\
  Swift/Rest & 24.3 &$ 2.16~10^{-5}$ & H1 \\
  \hline
\end{tabular}

\caption{$H_0$ is the null hypothesis : $\Delta\chi^2 < 16.27$ or
p-value $> 0.001$; $H_1$ is the alternative hypothesis.}
 \label{tab3}
\end{table}

From these results (Table 3), we conclude that the null hypothesis
is accepted for the \textit{BATSE} data. Therefore, the BATSE data
are well represented by two groups or populations of gamma-ray
bursts, as was originally suggested before by
\cite{kouveliotou:93}. On the contrary, for \textit{Swift/BAT}
data (both intrinsic and observed), the model of two groups (short
and long durations distributions) is clearly rejected by the
statistical test with a p-value less than $3~10^{-5}$.

\section{Discussion}

GRBs have widely been classified into two categories: short ones
(SGRBs) and long ones (LGRBs), separated at $T_{90}^{obs}$ = 2
sec, following a classification by \cite{kouveliotou:93} of
\textit{BATSE} GRBs. However, GRBs observed by different
instruments spread out into roughly log-normal distributions that
more or less overlap.

Our present work shows that the distribution of \textit{BATSE}
GRBs' observed durations can  be better fitted by two groups than
three groups; that is, the \textit{BATSE} bursts can be,
statistically, categorized into two classes better than three
classes. In some previous work \citep{{hakkila:00a},
{hakkila:00b}, {meegan:00}}, it has been reported that this
intermediate group is due to the selection effect of the
\textit{BATSE} instrument. And indeed, It has been therefore
difficult to present GRBs of intermediate durations as a
well-specified category \citep{bromberg:13}.

For \textit{Swift/BAT} observed durations, however, our
statistical analysis method shows a clear preference for a model
with three classes over one with two, with an intermediate class
appearing at durations in the range of 2 - 10 seconds. In fact, it
is worth noting that the intermediate class has a stronger
presence than the (traditional) short one, even if we limit the
fit to two groups. This was also true for the \textit{BeppoSAX}
bursts studied by \cite{horvath:09}.

For \textit{Swift/BAT} intrinsic durations, the results of our
study clearly shows a preference for the class of intermediate
duration over the short one, even in the two-group model. However,
the statistical test prefers the hypothesis of fitting the data
with three groups instead of two.

%%%%%%%%%%%%%%%%%
The classification of GRBs into two or three groups should in
principle be linked to intrinsic differences in the bursts, to
their mechanisms of production, and to their origins. SGRBs are
generally understood to be the result of a merging of two compact
objects (two neutron stars or one neutron star and one black hole)
and LGRBs are associated with collapses of hyper-massive stars
(``Collapsars"). A third group of only six GRBs, characterized by
a low luminosity ($L < 10^{49}~erg/s$), has been shown not to fit
in either of these two classes \citep{bromberg:13}. These authors
have also shown that the division at $T_{90}~\approx~ 2$ seconds
between Collapsars (long bursts) and non-Collapsars (short burst)
can no longer be ascertained, as 40 percent of Swift's bursts of
duration less than 2 seconds are in fact Collapsars; moreover, a
non-negligible fraction of those have durations less than 0.5
seconds; likewise, a number of non-Collapsars have durations
longer than 10 seconds. \cite{bromberg:13} suggest a division
between the two groups at 0.8 seconds as more suitable for
\textit{Swift/BAT}.

Our results also indicate a difference between the distributions
of durations presented by the \textit{BATSE} and
\textit{Swift/BAT} detecting instruments, which implies that the
detectors do significantly affect the categorization of bursts.
The relative dearth of SGRBs in the \textit{Swift/BAT} sample may
be due to the energy band (15 - 150 keV) in which the detector
operates, especially since SGRB photons are characterized by
relatively hard spectra. The same thing can be seen in the
distribution of GRBs from \textit{BeppoSAX}, which had operated in
the 40 - 700 keV energy band.

Finding a statistically significant intermediate class of bursts,
\cite{horvath:08} believe that the group is distinct and real; in
fact, they find distinctive physical features for these bursts: in
particular, 99.9 \% in logH43 and logH32 hardnesses and
anisotropic sky distribution \citep{{meszaros:00},{litvin:01}}.
These researchers also recall that this intermediate class had
been found in \textit{BATSE} bursts by a number of other
researchers, starting with \cite{mukherjee:98}.

On the other hand, one should note that finding evidence of a
third log-normal component in the analysis of the burst duration
distribution does not necessarily imply the existence of a third
physical channel for the production of GRBs, in addition to the
merger and collapsar mechanisms.  Indeed, the durations distribution
corresponding to the collapsar
scenario has no reason to be exactly symmetric in Log($T_{90}$).
If, for example, we link the duration of a burst to the amount of
mass that can be accreted by the newly formed black hole, the
durations distribution may reflect the distribution of envelope
masses of the progenitors. If this distribution is asymmetric, with
more massive progenitors being less common, a ``shoulder''
may appear in the durations distribution.  This shoulder would result
in a preferred three-group fit of the
data without imposing a new physical mechanism of burst formation.

\section{Conclusion}

In this work we set out to study the existence of a third,
intermediate class of bursts between the short-duration and
long-duration ones. The non-symmetric shape of the distribution of
long-duration GRBs seen in the \textit{Swift/BAT} data (a
``shoulder" on the lower side of the distribution) motivated us to
search for a separate group in the zone between 2 and 10 seconds.

Indeed, the $\chi^2$ statistical test  favors a distribution with
three groups in \textit{Swift/BAT} data, where the third class of
bursts appears rather clearly, contrary to the BATSE data.
Converting to intrinsic durations does not strongly modify the
general shape of the (\textit{Swift/BAT}) distribution.

The reality of a third, distinct class of bursts, with different
physical mechanisms, features, and sources, is far from
established. As we mentioned, besides the issue of detector
characteristics and their impact on the types of bursts that are
observed and recorded, the work of \cite{bromberg:13} has shown that
the two traditional classes (Non-Collapsars, i.e. Mergers,
producing ``short" bursts with durations less than 2 seconds, and
Collapsars, producing ``long" bursts with durations longer than 2
seconds) is no longer tenable. There are short Collapsars and long
Mergers. Perhaps the ``intermediate" class that our (and others')
statistical analyses indicate is simply underscoring this overlap
between the two types. One conclusion that seems inescapable is
that the division at 2 seconds is untenable.

\begin{acknowledgments}

 The authors gratefully acknowledge the use of the online \textit{Swift/BAT} table compiled by
 Taka Sakamoto and  Scott D. Barthelmy and the use of the online \textit{BATSE} current catalog presented by BATSE
 team. We acknowledge J.L. Atteia for useful discussions.  We thank the referee for very useful comments,
 which led to significant improvements of the paper.
\end{acknowledgments}
 %\section*{References}
%\bibliographystyle{alpha}
%\bibliographystyle{break}
%\bibliographystyle{abbrvnat}
%\bibliographystyle{ieeetr}
%\bibliographystyle{phaip}
%\bibliographystyle{plain}
%\bibliographystyle{unsrt}
%\bibliographystyle{apalike}
%\input{references.tex}
%\bibliographystyle{mn2e}
%\rhead{\textit{Bibliographie}}
\bibliographystyle{spr-mp-nameyear-cnd}
\bibliography{Astroph2}
\end{document}